\documentclass[12pt]{article}
\usepackage[utf8]{inputenc}
\usepackage{geometry}
\usepackage{mathptmx}
\usepackage{multicol}
\usepackage{changepage}
\usepackage{tabto}
\usepackage{authblk}
\usepackage{graphicx}
\usepackage{lettrine}
\usepackage{caption}
\usepackage{xcolor}
\usepackage{amsmath}

\usepackage{lipsum}
\usepackage{mathtools, nccmath}
\usepackage{braket}
\usepackage{ulem}

\rmfamily
\title{Decoherence puzzle in measurements of photons originating from electron-positron annihilation}
\author{Sushil Sharma${}^{a,b,c}$\thanks{sushil.sharma@uj.edu.pl} ,~Deepak Kumar$^{a,b,c}$, and Pawe{\l} Moskal$^{a,b,c}$ \\on behalf of the J-PET collaboration

{\it\normalsize $^a$Faculty of Physics, Astronomy, and Applied Computer Science, Jagiellonian University, Kraków, Poland}\\ 
{\it\normalsize $^a$Total-Body Jagiellonian-PET Laboratory, Jagiellonian University, Kraków, Poland}\\
{\it\normalsize $^a$Center for Theranostics, Jagiellonian University, Kraków, Poland}}
\date{}
\begin{document}
\maketitle
\begin{adjustwidth}{1.5cm}{1.5cm}
Entanglement of photons originating from the electron-positron annihilation has not been proven experimentally.~Though the independent experiments performed so far unanimously confirm that correlation between the linear polarizations of back-to-back photons from the electron-positron annihilation is consistent with the assumption that these photons are entangled in polarization.~Yet, unexpectedly, recent experiments differ as regards the correlation of polarization direction of back-to-back photons after the decoherence induced by scattering of one of these photons on the electron in the scattering material. In one of the experiments, the correlation before and after the decoherence of the photon state is the same and in the other experiment the scattering of one photon leads to a significant decrease in this correlation. Here we discuss this puzzle. The decoherent states were ensured by forcing one of the annihilation photons through prior scattering  before the Compton kinematics-based measurement of the polarization correlation. A comparison between the experimental setups used for the different measurements and the results obtained are briefly discussed, highlighting the parameters that are important for performing such measurements. Finally, the main features of the J-PET detector are presented with the schemes for performing similar measurements so that the results can be used as conclusive remarks for solving this puzzle. Solving the decoherence puzzle will have crucial consequences for basic studies of entanglement, as well as for the proposed application of polarisation of photons in positron emission tomography.  In case if the correlation of polarisation of back-to-back photons from the electron-positron annihilation is the same before and after the scattering of these photons, then it will not be useful for the reduction of the scatter fraction in PET diagnostics.\\\\
\textbf{keywords}: Quantum entanglement, Positron Emission Tomography (PET), Compton polarimeters, plastic scintillators
\end{adjustwidth}
\section{Introduction}
\begin{multicols*}{2}
\lettrine{U}{}nderstanding the entangled property of \mbox{photons} originating in $e^+e^-$ annihilations is crucial not only for the fundamental study of the quantum behavior of bound system of purely leptonic objects ( $e^-$ and its antiparticle $e^+$), but also because of its direct implications for PET modalities. The idea was born as early as 1930 with Dirac's article “On the Annihilation of electrons and protons", in which he discussed the emission of two photons in annihilation processes, which became the basis of the well-known pair theory~\cite{Dirac1930}. Based on the pair theory, J.A. Wheeler conceptualized the polarization correlation of photons emitted from $e^+e^-$ system in opposite directions~\cite{WHE46}. He stated that if one of the two annihilation photons is linearly polarized in one plane, then the other photon with the same momentum but in opposite direction will be linearly polarized in the perpendicular plane. To confirm this hypothesis, Wheeler proposed an experiment in which a slow positron interacts with an electron at rest to produce two photons. This was relied on measuring the azimuthal correlation between the polarization of two photons based on Compton scattering as a polarization analyzer for high-energy 511 keV photons.~According to the Klein-Nishina \mbox{formula~\cite{NIS29}}, a photon will most likely scatter perpendicular to the direction of its linear polarization at a random angle of the azimuthal plane. Similarly, the other photon will also scatter at a preferred azimuthal angle. Thus, Compton scattering of both photons can result in preferential registration of the photons with the chosen polarization. Therefore, the polarization correlation can be measured by calculating the ratio when the relative azimuthal angles of both photons are $\pm$90$^\circ$ (they are scattered perpendicular to each other, N$_\perp$) to 0$^\circ$ (both photons are scattered parallel or antiparallel, N$_\parallel$).~Wheeler also predicted that maximum value of this ratio would be expected at scattering angles of 74.30$^\circ$. The next year, Ward noticed an error in Wheeler's prediction of the two-photon wave function of entangled photons - the \mbox{interference} term was neglected. Ward corrected it, claiming that the maximum ratio (N$_\perp$/N$_\parallel$) would be at scattering angles of 82$^\circ$. Inspired by the scheme described by Wheeler, Ward proposed a scheme for the experimental setup for performing the measurement~\cite{WARD47}. He also derived the double differential cross section for the scattering of two linearly polarized photons by $\theta_1$ and $\theta_2$ at the respective azimuthal angles $\phi_1$ and $\phi_2$~\cite{WARD47}. This is expressed as:
\begin{equation}\label{DDx}
\begin{aligned}[b]
 \frac{d^{2}\sigma(\theta,\phi)}{d\Omega_{1}d\Omega_{2}}= &\frac{r_{e}^{4}}{16}\left[A(\theta_{1},\theta_{2})-B(\theta_{1},\theta_{2})cos(2(\Delta\phi))\right]
\end{aligned}
\end{equation}
where
 \[A(\theta_{1},\theta_{2})= \frac{\{(1-cos\theta_{1})^{3}+2\}\{(1-cos\theta_{2})^{3}+ 2\}}{(2-cos\theta_{1})^{3}(2-cos\theta_{2})^{3}}\]
 \[B(\theta_{1},\theta_{2})= \frac{sin^{2}\theta_{1}sin^{2}\theta_{2}}{(2-cos\theta_{1})^{2}(2-cos\theta_{2})^{2}}\]
$\theta_1$ and $\theta_2$ are the scattering angles, $\Delta\phi$ is the relative azimuthal angle ($\phi_1 - \phi_2$) for two photons and r$_e$ is the electron radius. It is worth noting that a similar equation was also derived independently by Snyder et al.~\cite{SNY48}.\\
Following the proposed experimental scheme, two measurements were made in the following year by Bleuler et al.~\cite{BLE48} and Hanna~\cite{HAN48}.~Both experiments observed the correlation between the linear polarizations of two photons using Compton kinematics.~The correlation ratio obtained in both measurements was consistently lower than the theoretically predicted value (eq.\ref{DDx}). However, Hanna pointed out some important observations that could account for the discrepancies between theoretically predicted and experimentally measured values, eg., (i) the absence of low-density scatterers, (ii) lack of an efficient gamma-ray counter and (iii) limited geometric acceptability for studying events of interest. In 1950, Wu and Shaknov repeated the experiment, but with an improved aluminium scatterer (0.5 inch diameter, 1 inch length) and newly developed gamma-ray detectors based on anthracene crystals coupled to RCA 5819 photo-multiplier tubes~\cite{WU50}. The anthracene-based scintillator counters had 10 times better efficiency than the Geiger counters used in previous experiments~\cite{BLE48,HAN48}. Wu et al.~\cite{WU50} successfully obtained experimental results with a correlation ratio of 2.04 $\pm$0.08, which was consistent with calculated values of 2.0 when geometric acceptance was considered~\cite{WU50}. It should be emphasised that all of these experiments differed in their experimental configuration and required modification of the Pryrc-Ward formula~\cite{WARD47} to account for corrections for geometric effects before comparison with the experimental results. Later, two more experiments were performed to measure the correlation ratio. The first of these studies was conducted by Langhoff in 1960.~He \mbox{reported} the results of thorough measurements with improved geometry by measuring the correlation ratio at various azimuthal angles. However, a good agreement was obtained between the estimated \mbox{theoretical} prediction (2.48$\pm$0.02) and experimentally measured value (2.47$\pm$0.07) for polar scattering angle $\theta$=82$^\circ$~\cite{LAN60}. The second study was performed by Kasday et al~\cite{KAS75}, in which they explicitly applied several sources of correction that could be necessary for the correct estimation of such a correlation. They propose to rewrite the eq.~\ref{DDx} as follows:
\begin{equation}\label{DDx22}
\begin{aligned}
 P(\Delta \phi)&=\frac{r_e^4A(\theta_1,\theta_2)}{16}\left (1- \frac{B(\theta_1,\theta_2)}{A(\theta_1,\theta_2)}cos(2\Delta \phi) \right)&\\
 P(\Delta \phi) &= k(1-\nu cos(2\Delta \phi))&
\end{aligned}
\end{equation}
where k=$r_e^{4}$A($\theta_1$,$\theta_2$)/16 and $\nu$ = B($\theta_1$,$\theta_2$)/A($\theta_1$,$\theta_2$) are functions sensitive to the scattering angles. To measure the correlation of linear polarization of annihilation photons, the correlation ratio (R) can be calculated when $\Delta\phi=\pm90^\circ$ and $\Delta\phi=0^\circ$ and the expression for R is defined as: 
\begin{equation}\label{DDx2}
\begin{aligned}[b]
 R(\Delta \phi )=\frac{P(\Delta \phi=\pm 90^\circ)}{P(\Delta \phi=0 )}=\frac{1+\nu}{1-\nu}
\end{aligned}
\end{equation}
which has the maximum value of  2.85 at $\theta_1$=$\theta_2$=82$^\circ$~\cite{WARD47,SNY48}.\\
\par With the possibility of experimentally observing the correlation between the linear polarization of the annihilation photons, which agrees with the theoretically predicted values, Bohm and Aharnov pointed out that this can be considered as a similar case of quantum entanglement as discussed by Einstein, Podolsky and Rosen (EPR) ~\cite{EIN35}. Following the experimental setup of Wu et al ~\cite{WU50}, Bohm and Aharnov derived results that allowed using the values of R ~\cite{BOH57} to establish upper limits for two hypothesis: (i) Entangled state (of 2 photons linearly polarized in (x$_1$,y$_1$), (x$_2$,y$_2$), respectively) can be expressed by the wave function~\cite{WARD47}: $\psi=1/\sqrt{2} (\Ket{x}_1\Ket{y}_2 - \Ket{y}_1\Ket{x}_2)$. (ii) hypothetical separable state (wave function does not overlap, but polarizations are orthogonally correlated). An upper limit of 2.85 was calculated for the entangled state, while a value less than 2 was predicted for the above mentioned separable state . Shortly thereafter, Wilson et al.~\cite{WIL76} reported on the influence of the distance between the Compton polarimeters and the origin of the annihilation photons in measuring the correlation ratio (R), where no significant change was observed even at a distance of 2.5~m. Therefore, such measured polarization correlation between photons emitted from $e^+e^-$ in the back-to-back direction and forming the Line Of Response (LOR) for imaging the annihilation source can be used to reduce unwanted coincidences accepted during imaging~\cite{MCN14,TOG16}. In a recent study by Watts et al.~\cite{WAT21}, the use of the entangled polarization correlation to distinguish true events (no scattering before detection, entangled state) and scattered events (at least one of the two photons is prior scattered before registration in the scanner and thus defined a case of decoherence, i.e., loss of entanglement) was demonstrated. The authors showed that in the latter case the correlation between the relative polarization amplitudes was much lower than in the former one. Thus, by selecting only those events for which the relative polarization is entangled, one can significantly improve the image quality by suppressing the background. Moreover, they also determined the exact value of the upper limit for the separable state of 1.63 by introducing the correction into the formula derived in the Bohm paper (supplementary note in DP Watts et al.~\cite{WAT21}).\\
In contrast, Abdurshitov et al~\cite{ABD22}, reported that regardless of the initial state (entangled or decoherent) of the annihilation photons, the same polarization correlation distribution was measured. This left open the question of whether the measured correlation between the polarization of annihilation photons can be considered a unique signature for distinguishing the origin of annihilation photons from entangled or separable states. In both studies, a different detector scheme was used to measure the correlation ratio based on Compton kinematics. However, the methodology used to define the entangled and non-entangled (decoherent) states were nearly identical. In Sec.~\ref{PreResults}, the experimental schemes of the two experiments are briefly discussed and their results are compared. The main goal of the present work is to address this entanglement puzzle in measuring the relative polarization of photons. Also, the J-PET tomograph is described, as well as how it can be used as a potential detector to measure the relative correlation between the linear polarization of annihilation photons in order to find a conclusive solution to the entanglement puzzle~\cite{MOS22,NOW17}.\\ 
\section{Recent results of experimental measurements leading towards the puzzle in~{witnessing} entanglement}\label{PreResults}
In work of Watts et al~\cite{WAT21}, the hypothesis proposed by Bohm and Aharonov~\cite{BOH57} for entangled and separable states was tested experimentally. For the experiment, two cadmium-zinc-telluride (CZT) detectors (each detector was a 1~cm cube divided into 121 pixels of size 0.8x0.8~mm$^2$) were placed at 8.7 cm apart. A $^{22}$Na source with an activity of 170 kBq emitting 511 keV photons in opposite directions, housed in a plastic case, was placed in the center of two detection modules. The measurements were performed with two different experimental setups. In the first setup (Fig.~\ref{PetDemons}(a)), the detectors were aligned along the axial alignment to the source to register annihilation photons (511 keV) and the corresponding scattered photons to test the entanglement hypothesis~\cite{WARD47,SNY48,BOH57}. To test the second hypothesis, i.e., the effect on the value of the correlation ratio (R) when one of the two annihilation photons is forced to scatter before it interacts in the CZT detector. Prior scattering of one of the entangled photon pairs results in a loss of entanglement. To perform this experiment, one of the CZT detectors was rotated by 33$^\circ$ and a scattering medium (nylon) was placed along the path of one of the photons on the same side so that a scattered photon would interact with the rotated detector to determine the correlation ratio (see Fig.~\ref{PetDemons}(b)).
For the first case, the measured value of the correlation ratio R = 1.85$\pm$0.04 for the selected scattering angle range was 70$^\circ$-110$^\circ$, still less than the theoretically predicted value for entanglement (2.85), but more than the theoretically predicted upper limit for separable states~\cite{BOH57,WAT21}. The measured distribution ($\Delta \phi$) of the data was well described by the QE-PET simulation results. QE-PET was developed based on the Geant4 toolkit, which incorporates the entanglement formalism for the primary interaction of annihilation photons with detectors instead of the standard Klein-Nishina formula for polarized photons; see ~\cite{WAT21} for details.
\begin{center}
\includegraphics*[width=\linewidth]{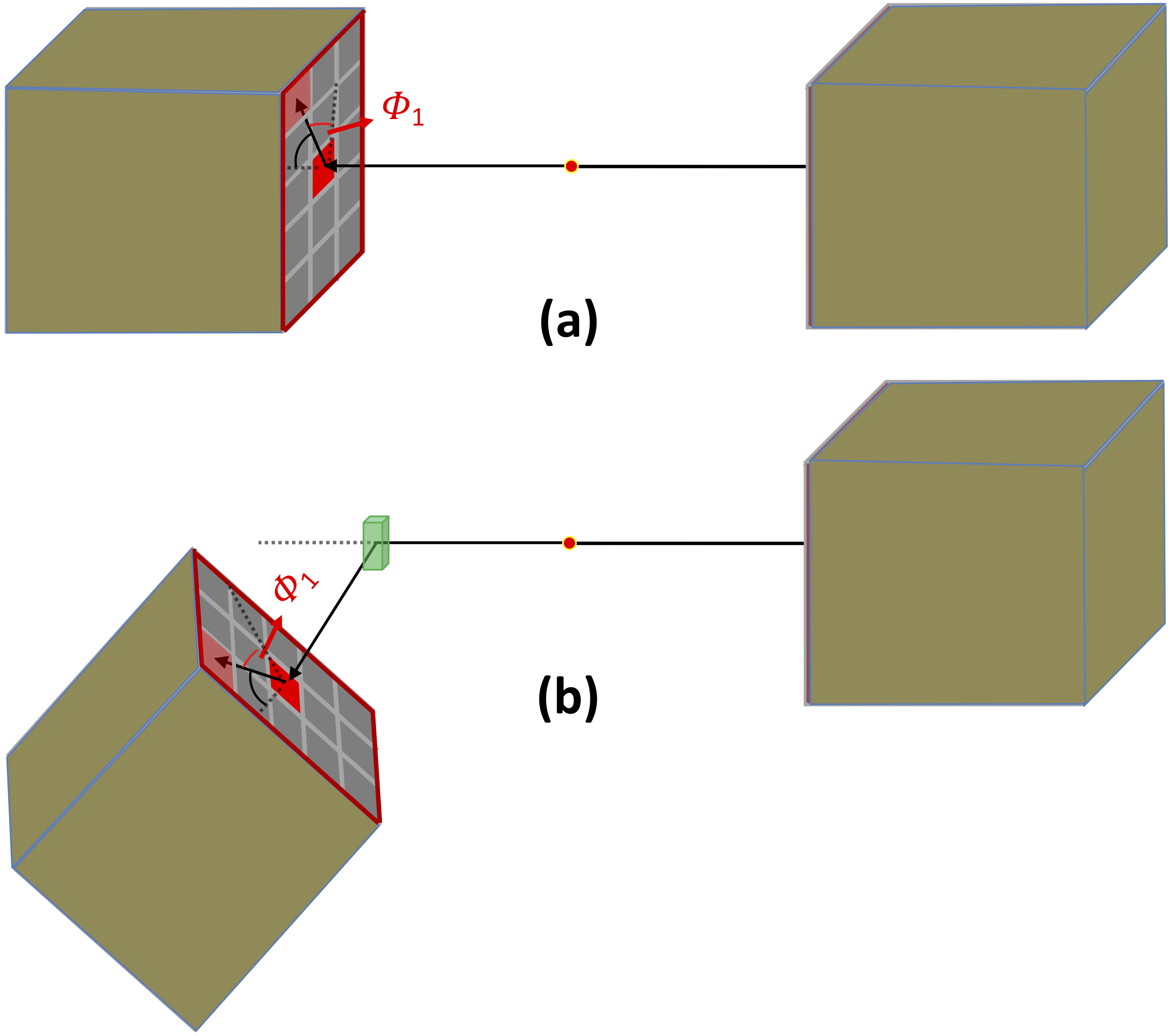}    
\captionof{figure}{\label{PetDemons}Schematic diagram of the measurement schemes used in reference ~\cite{WAT21}. A cube-shaped detector with each side of 1~cm was divided into 121 pixels to read the signals. (a) Shows the arrangement of two detectors to test the hypothesis that annihilation photons emit in opposite directions and exhibit entangled polarization. (b) Shows the experimental setup for measuring the correlation ratio for decoherent state, when one of the photons was scattered by the nylon scatterers and the lose of entanglement between the linear polarizations of the photons was assumed. For this purpose, one of the detectors (left) was rotated by an angle euqal to 33$^\circ$.}
\end{center}
With more narrow range of scattering angles (93$^\circ$-103$^\circ$), the value of R reached 1.95$\pm$0.07. \\\\
However, significant suppression was observed for the decoherent state (in both measured and simulated results). Given the large statistical uncertainty, which could be due to the small aperture angle for the detection of the scattered photons, it is difficult to draw firm conclusions as to whether the measured relative correlation was suppressed by a hypothesis-based formalism~\cite{BOH57} or whether more advanced measurements with broad geometric acceptance were required. In summary, it was reported that the kinematics of Compton scattering in orthogonally entangled annihilation photons is different from that in which entanglement is considered lost because of the prior scattering of one of the photons. It is worth mentioning that the obtained results~\cite{WAT21} do not agree with the theoretical predictions~\cite{ BEA19}.
\begin{center}
\includegraphics[height=0.72\linewidth]{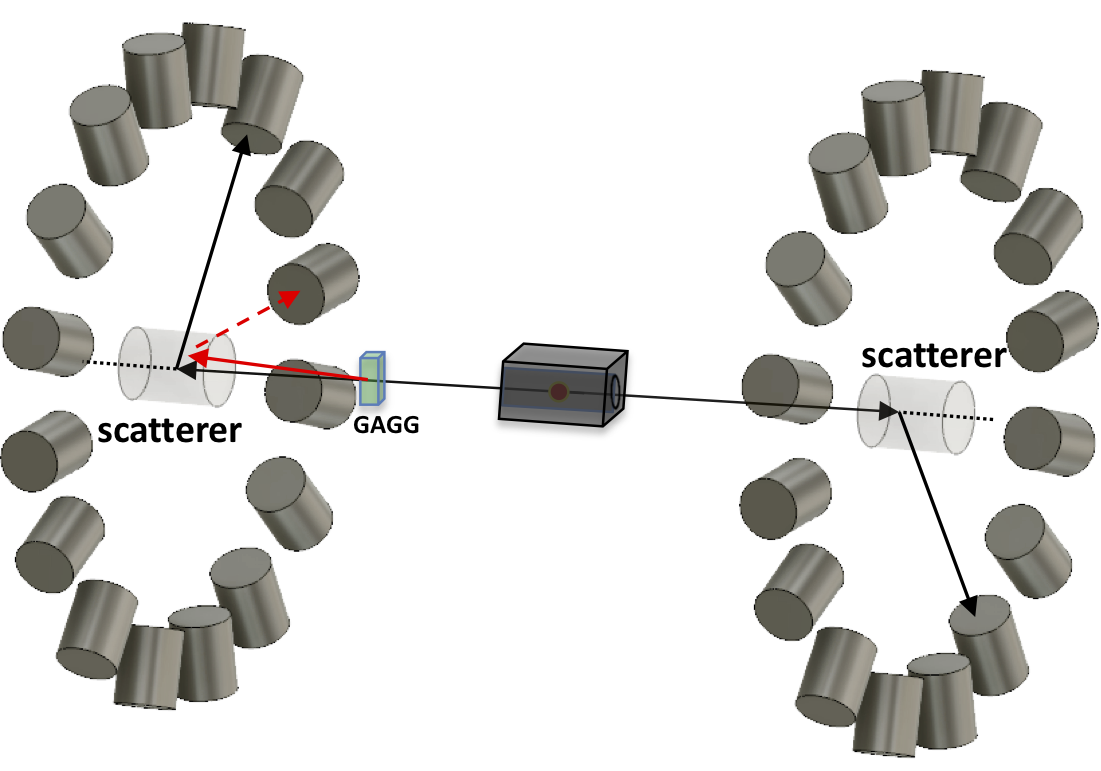}
\captionof{figure}{\label{Polari}Schematic of the experimental setup used by Abdurashitov et al.~\cite{ABD22}. Two Compton polarimeters were used, each consisting of 16 NaI(Tl) detectors. The red dot in the center represents the placement of the source within the perforated lead collimator. The transverse cylinders in the center of each polarimeter are the plastic scatterers, while the green scatterer is a GAGG scintillator with a separate signal readout used primarily to induce decoherence in the polarization correlation of the annihilated photons. The red solid and dashed lines represent the case where a photon passing through the GAGG is scattered.}   
\end{center}
Recently, Abdurashitov et al.~\cite{ABD22}, reported experimental results that contradicted those of obtained by Watts et al~\cite{WAT21}. The experiments were performed using the same methodology: (a) when the photons relative linear polarization was considered to be entangled (direct Compton kinematics applied to back-to-back photons to measure the correlation), (b) inducing decoherence in the polarization entanglement, using the scatterer along the path of one of the photons, and calculating the effect on the measured correlation. The experiments were performed with a pair of Compton polarimeters.
The scheme of the experiment is shown in Fig.~\ref{Polari}. Two Compton polarimeters, spaced about 70 cm apart and each consisting of 16 NaI(Tl) detectors as photons counters of size 5x5 cm$^2$, were aligned with the plastic scatterer (transparent cylinder) in the center, which was equidistant from all NaI crystals. To produce the annihilation photons, a $\beta^+$ emitter source ($^{22}$Na) with an activity of 50~MBq was placed in a lead shield with a perforated cylindrical collimator. To induce decoherence in the polarizations, a gadolinium-aluminum-gallium-garnet (GAGG) was placed along the path of one of the annihilation photons. Events without and with the interaction of the photon inside the GAGG (when the energy deposition was in the range of 10-40 keV, considered as interaction) were characterized as with entangled and decoherent polarizations, respectively. In the proposed experimental scheme, the annihilation photons interacted with the respective scatterers and scattered photons were registered in the NaI(Tl) counters of the polarimeters. The analysing power of the Compton polarimeter can be estimated as: $A(\theta)=(N_\perp - N_\parallel)/(N_\perp + N_\parallel)$. 
The differential cross section for Compton scattering of a linearly polarized photon can be calculated using the Klein-Nishina formula~\cite{NIS29}:
\begin{equation}\label{KNis}
\begin{aligned}[b]
&\frac{d\sigma }{d\Omega }=r_{e}^{2}.\epsilon^{2}(\epsilon+\frac{1}{\epsilon}-2sin^{2}\theta cos^{2}\phi )\\
&= r_{e}^{2}\epsilon^{2}(\epsilon +\frac{1}{\epsilon }-sin^{2}\theta )(1-\frac{sin^{2}\theta }{\epsilon +\frac{1}{\epsilon }-sin^{2}\theta }cos(2\phi ))\\
&=r_{e}^{2}k(1-\alpha(\theta )cos(2\phi ) )
\end{aligned}
\end{equation}
where $\epsilon$ is the ratio of E$^{'}$ and E, the ratio of scattered to incident photon energies. $\theta$ is the scattering angle and $\phi$ is the azimuthal angle (angle between the scattering plane and the direction of linear polarization of the incident photon). k is equal to $\epsilon^{2}(\epsilon +1/\epsilon-sin^{2}\theta)$ and $\alpha(\theta) = sin^{2}\theta /(\epsilon + 1/\epsilon -sin^{2}\theta)$. Analyzing power of the Compton polarimeter can be described as~\cite{KNI18}:
\begin{equation}
\begin{aligned}[b]
 &   A(E,\theta)=\frac{\frac{d\sigma }{d\Omega }(\theta ,\phi =90^\circ)-\frac{d\sigma }{d\Omega }(\theta ,\phi =0^\circ)}{\frac{d\sigma }{d\Omega }(\theta ,\phi =90^\circ)+\frac{d\sigma }{d\Omega }(\theta ,\phi =0^\circ)}\\
 & \hspace{1.24cm}  =\frac{sin^{2}\theta}{\frac{E^{'}}{E}+\frac{E}{E^{'}}-sin^{2}\theta }  = \alpha(\theta)
    \end{aligned}
\end{equation}
For a photon of energy 511 keV scattering at $\theta=82^\circ$ the value of A = $\alpha(\theta)$ reaches a maximum (A=0.69). The probability of Compton scattering of two orthogonally polarized photons scattering at angles $\theta_1$ and $\theta_2$ can be written in terms of the analyzing power of Compton polarimeters as follows~\cite{ABD22}:
\begin{equation}
P (E_1,E_2,\Delta \phi)=r_{e}^{2}k_1k_2(1-\alpha(\theta_1)\alpha(\theta_2)cos(2\Delta \phi ) )
\end{equation}
Finally, a polarization modulation factor ($\mu$) estimating the relative polarization of annihilation photons can be calculated as follows~\cite{BEA19,CAR19}:
\begin{equation}
\mu =\frac{P(\Delta \phi=90^\circ)-P(\Delta \phi=0)}{P(\Delta \phi= 90^\circ)+P(\Delta \phi =0)} = \alpha(\theta_1)\alpha(\theta_2))
\end{equation}
The modulation factor is equal to the product of the analyzing powers of the individual polarimeter. For the scattering angles $\theta_1$=$\theta_2$=82$^\circ$, the modulation factor is equal to 0.48. In the comparative calculation to the correlation ratio (R), $\mu$ is equivalent to $\nu$ as used in eq.~\ref{DDx2}. The experimentally determined value of $\mu$ was 0.41, which corresponds to R$\approx$2.39 and proved to be in better approximation to the theoretically predicted value (R=2.85). 
%
\begin{center}
\includegraphics*[width=0.8\linewidth]{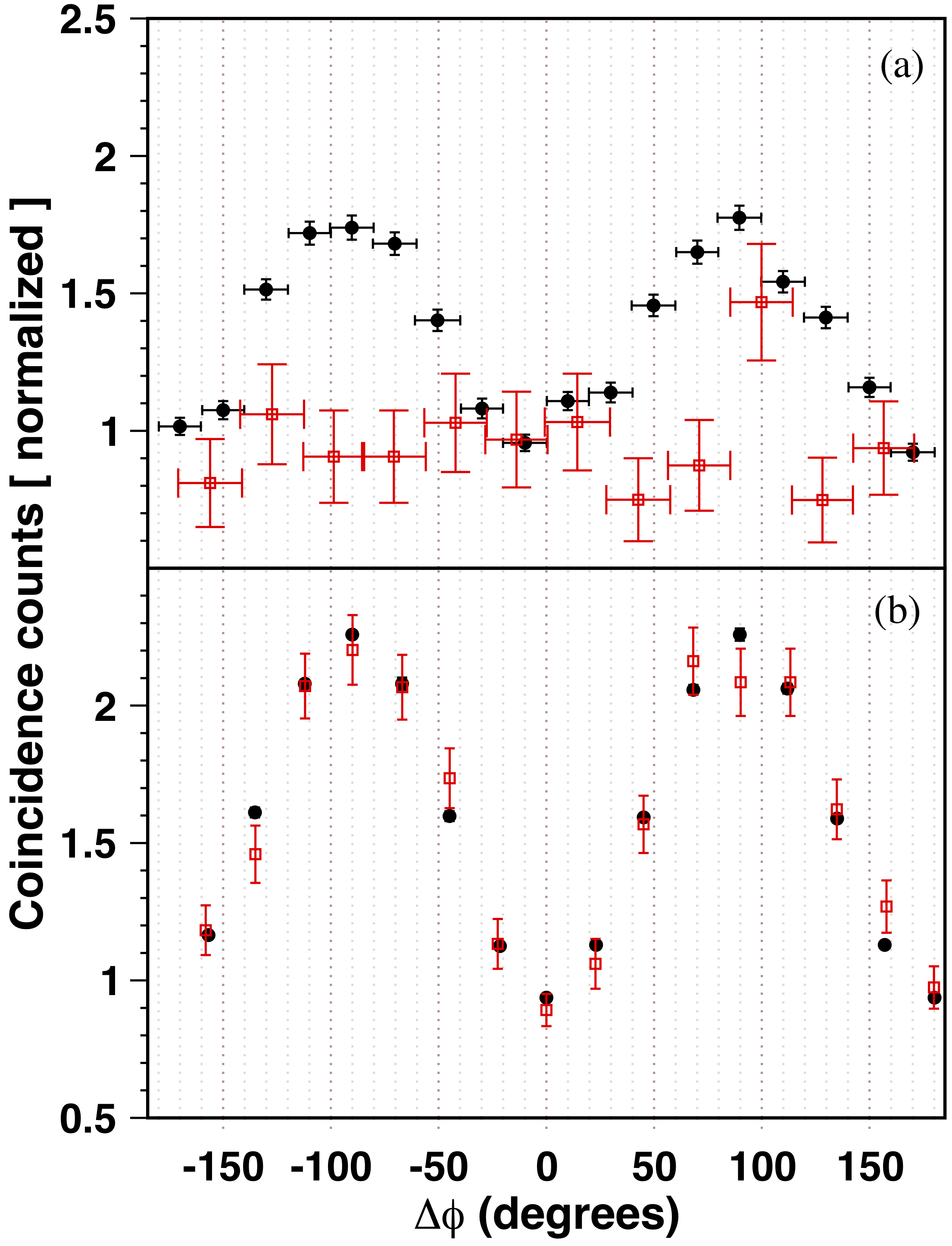}
\captionof{figure}{\label{ResultsComp} The results of measured polarization correlation of annihilation photons for entangled (black filled circles) and decoherent (red squares) states. (a) The panel shows the results of ~\cite{WAT21}. For the data presented here, a scattering angle range of 70$^\circ < \theta_1=\theta_2 < 110^\circ$ was chosen to investigate the entangled state hypothesis, while 60$^\circ < \theta_1=\theta_2 < 140^\circ$ was chosen for the decoherent state. (b) The panel shows the results of ~\cite{ABD22} for two different states. In this study, the scattering angle range was 80$^\circ < \theta_1=\theta_2 < 100^\circ$ for both cases. Details of the experiments can be found in the respective publications~\cite{WAT21,ABD22}.}    
\end{center}
It should be emphasized that in the work of Abdurashitov et al.~\cite{ABD22}, no difference in the $\mu$ distribution was observed for the entangled and decoherent quantum states, indicating that the Compton kinematics remained identical for both cases as predicted in~\cite{BEA19}. Fig.~\ref{ResultsComp} shows the results of the measured correlation for entangled (black filled circles) and decoherent states (red squares) reported in both publications. 
Panel (a) shows the results of~\cite{WAT21} and panel (b) shows the results of~\cite{ABD22}. In both cases, only experimental data from the measurements are shown. For comparison, the distributions of $\Delta \phi$ from ~\cite{ABD22}, were translated into the azimuth distribution range -180$^\circ$ to 180$^\circ$. In addition, the results of ~\cite{ABD22} were normalized in the same way as proposed in ~\cite{WAT21} by normalizing the respective distributions with average yields for $\Delta \phi=0,\pm180^\circ$. In both papers, using an experimental setup to measure annihilation photons with entangled relative polarization, an enhancement of the measured relative polarizations was observed. Similar results have been reported by other groups~\cite{KOZ19,PAR22}. However, the contradictory results in the measurement of this correlation for the decoherent states (induced entanglement loss) require further attention. For this particular case, the results reported in~\cite{WAT21} are subject to large uncertainties. In~\cite{ABD22}, on the other hand, the threshold applied to distinguish between the entangled and decoherent cases (based on energy loss criteria (10-40 keV) with prior scattering) may not be sufficient for entanglement loss or to observe the difference in correlation between these two cases. Therefore, to solve this puzzle, a more detailed study is needed, especially with a detector that can overcome to some extent the possible bias due to experimental limitations. In this context, we propose the new measurement with the J-PET detector, the first tomograph based on plastic scintillators. The details of the J-PET detector and the scheme for making such measurements are described in the next section.
\section{JPET as Compton polarimeter to measure relative polarization}
The \textbf{J}agiellonian \textbf{P}ositron \textbf{E}mission \textbf{T}omograph (\textbf{J-PET }) is the first tomograph based on the idea of using long strips of plastic scintillators instead of crystal scintillators~\cite{MOS11,MOS16FF,MOS21F}. Plastic scintillators are mainly composed of hydrocarbons, so the predominant medium for the interaction of photons within the scintillator is Compton scattering~\cite{NIS29}. J-PET was designed and constructed as a multi-photon positron emission tomography scanner~\cite{MOS20FF,MOS21A}  capable of (i) standard PET imaging~\cite{MOS21F,SHO21F}, and (ii) newly invented positronium imaging~\cite{MOS21B,MOS19F,MOS19FF,MOS20F,MOS22F,MOS18,MOS16F}. For the planned measurement, we propose to use the 3-layer prototype of J-PET, which consists of plastic scintillators of length $50\times1.9\times0.7~cm^3$ read out at both ends by vacuum photomultipliers (see Fig.~\ref{ExperimentSetUp}). It is made-up of 192 plastic scintillators arranged in three concentric cylindrical layers with diameters of 85~cm, 93.5~cm, and 115~cm, respectively. The hit position along the length of the scintillator can be calculated as the measured time difference of the incoming light signals at both ends of the scintillator times speed of light divided by 2. The hit time, on the other hand, was calculated as the average of the measured times of the light signals~\cite{MOS14}. In J-PET, the data are stored in triggerless mode~\cite{PAL17,KOR18} and Time Over Threshold (TOT) is approached as measure of energy depositions~\cite{SHA20A,SHA20B}. In addition, a special framework was developed to analyse the stored data~\cite{KRZ20}. Since J-PET has a large coverage in both polar and azimuthal angles (see Fig.~\ref{ExperimentSetUp} and \ref{Entan}) and it offers an excellent angular resolution $\approx1^\circ$, it can be a potential detector to register the annihilation photons and their corresponding scattered photons. In Fig.~\ref{Entan} at the top, there is a picture depicting a single scintillator connected to photomultipliers. Below is the photo representing the 3-layer prototype that has been put into operation in the laboratory. The first and second layers consist of 48 plastic scintillators, while the third layer consists of 96 scintillators~\cite{NIE17F}.
\begin{center}
\includegraphics[width=\linewidth]{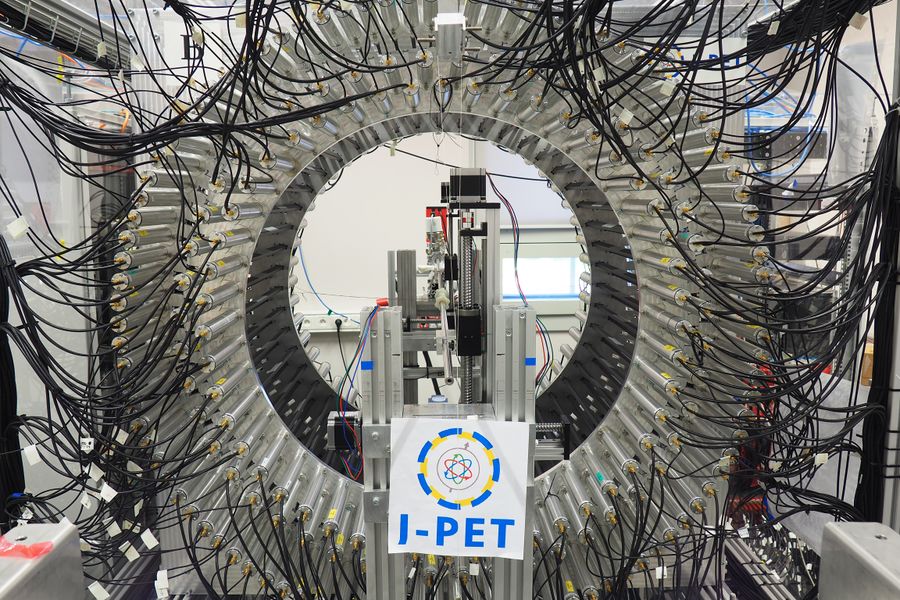}
\captionof{figure}{\label{ExperimentSetUp} Shows the 3-layer prototype of the J-PET detector currently operating in the laboratory.}    
\end{center}
 The scheme for measuring the azimuthal correlations between the polarizations of the annihilation photons based on their scattering planes is demonstrated. Here we would like to briefly discuss how the J-PET can be used in performing the experiment to study the polarization correlation for both cases (entanglement, decoherent) as described above:\\
\noindent \textbf{Entangled state:} Events with 4 hits will be examined. Of the four hits, two will be from 511~keV photons emitted in the theoretically assumed entangled polarization. The remaining two hits will be caused by the corresponding scattered photons (see Fig.~\ref{Entan}). Knowing the direction of the incident photon and the scattered photon, the polarization direction of the incident photon can be obtained from the measured hit positions: $\vec{\varepsilon _{i}}=\vec{k_{i}}\times \vec{k_{i}^{{}'}}$~\cite{MOS18}. The angular correlation between the polarization directions of the annihilation photons will be compared with theoretically predicted values~\cite{WARD47,BOH57}.\\
\noindent \textbf{Decoherent state:} To study this case, it is proposed that one of the photons undergoes prior scattering (to achieve entanglement loss ) before the polarization correlation is measured based on Compton scattering~\cite{WAT21,ABD22}. To our knowledge, there are only two studies~\cite{WAT21,ABD22} in which the correlation for decoherent states has been measured experimentally and compared with theoretically predicted values~\cite{BOH57}. In the case of J-PET, we propose to use the scintillators of the entire first layer as scatterers for the prior scattering to decohere at least one of the photons (5-hit events) or even both (6-hit events) to finally determine how the correlation changes in the case of an entangled and decoherent state.
\begin{center}
\includegraphics[width=\linewidth]{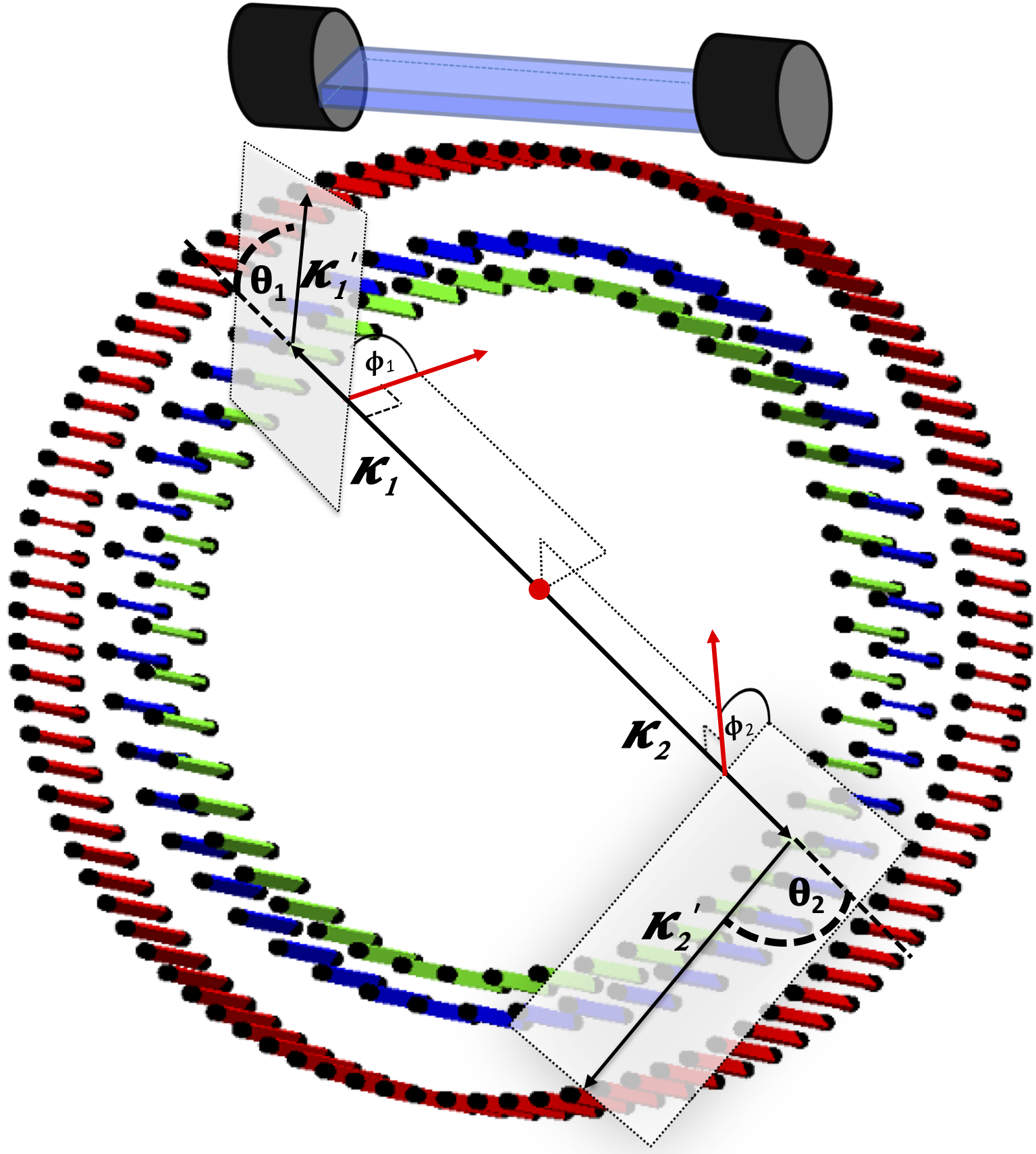}
\captionof{figure}{\label{Entan}J- PET detector made of 192 plastic scintillators read out by photomultipliers at each end is shown. A single detection module is also shown above. The red arrows represent the orthogonal polarization of annihilated photons emitted from the $e^+e^-$ system in opposite directions with momentum vectors \textbf{$k_{1}$} and \textbf{$k_{2}$}. The dotted planes on each side show the scattering planes of the photons, which are composed of the incident and the respective scattered photons. The scattering angles of annihilation photons are represented by $\theta_1$ and $\theta_2$, whereas the azimuthal angles between the scattered photon and the corresponding linear polarization are $\phi_1$ and $\phi_2$, respectively. The azimuthal correlation between the scattering planes can be calculated as $\phi_1$ - $\phi_2$.}    
\end{center}
The advantage of J-PET geometry is that the main process of photon interaction is Compton scattering. However, the assignment of the scattered photon to its primary origin, which is necessary to measure the scattering angle, is not trivial. This can be done using the scattering test, which uses the measured hit time and hit position of the interactions of primary and scattered photons in two different scintillators~\cite{SHA20B}. The scattering test is defined as $S=(t_{2}-t_{1})-D_{12}/c$, where $t_2$ and $t_1$ are the registration times of the scattered and primary photons, respectively, and $D_{12}$ is the distance calculated from the measured hit positions. In recent years, algorithms for studying positronium decays with the J-PET detector~\cite{DUL21} have been developed for research in fundamental physics~\cite{MOS21A} and medical physics~\cite{MOS21B}. For the various studies, a large amount of experimental data was stored by placing a $^{22}$Na source encapsulated in Kapton film and surrounded by porous material in a plastic chamber. These data can be used to perform the studies discussed in this article. Here, we only wanted to discuss the status of the current studies and the contradictory results in the measurement of the relative polarization of the annihilation photons by Compton kinematics, in particular, whether the quantum state of the Ps atoms was entangled or not before annihilation according to the theoretical predictions~\cite{BOH57}. The results of the detailed analysis of the data measured with J-PET are the subject of a detailed article, which we will report shortly.
\section{Conclusion and perspectives}
The measurement of orthogonally correlated linear polarization of photons emitted upon annihilation of the $e^+e^-$ system has regained interest due to its direct application in medical imaging. In 1957, Bohm and Aharonov suggested that such a measured correlation could be a signature of the entangled or non-entangled state of the $e^+e^-$ system prior to annihilation~\cite{BOH57}. Therefore, it is of utmost importance to measure this correlation accurately and to verify the theoretical predictions. It is suggested~\cite{BOH57,WAT21,ABD22} that this correlation can be measured for two different cases: (a) when the orthogonally correlated linear polarization of the annihilation photons is considered to be entangled and their relative polarization can be measured using Compton scattering as a tool for the polarization analyzer, (b) when the correlation is measured by the same method except that one of the annihilation photons has been previously scattered and this can be considered to be the case of non-entangled (decoherent) polarization of the annihilation photons. In such measurements, the proper choice of photon scatterers, detectors, and especially geometric coverage is of paramount importance to studying the Compton kinematics of such events. The first case was studied by several groups~\cite{WAT21,ABD22,KOZ19,PAR22} and all of them observed an enhanced correlation in the relative polarization of the photons. For the second case, the first experiment was performed by Watts et al~\cite{WAT21}. The results stated that no significant correlation was observed, however, the results obtained were reported with large uncertainties. In addition, the experimental results did not agree with the theoretical work~\cite{BEA19}. Abdurashitov et al~\cite{ABD22} also reported the experimental results, but their results contradicted the observations claimed by Watts et al~\cite{WAT21}. This raises questions and requires further investigation. Solving the puzzle of entanglement observation based on Compton kinematics of orthogonally polarized annihilation photons emitted in an $e^+e^-$ system is not only important to understand the fundamental process but also of outstanding importance because of its application in medical imaging. We propose to perform similar studies using the J-PET detector. The J-PET detector has demonstrated its potential in performing studies to register the decays of Ps atoms~\cite{MOS21B,DUL21}. Due to its geometric advantage, J-PET can study not only the two cases previously studied, but also the case in which both photons are scattered beforehand (instead of only one), since in these cases the correlation is measured and the entanglement loss can be assumed with greater certainty.
\vspace{-0.5cm}
\section{Acknowledgement}
The authors gratefully acknowledge the support of the Foundation for Polish Science through programme TEAM POIR.04.04.00-00-4204/17; the National Science Centre of Poland through grant nos. 2019/35/B/ST2/03562, 2021/42/A/ST2/00423 and 2021/43/B/ST2/02150; the Ministry of Education and Science under the grant No. SPUB/SP/530054/2022; EU Horizon 2020 research and innovation programme, STRONG-2020 project, under grant agreement No 824093; the Jagiellonian University via the project CRP/0641.221.2020, and via SciMat and qLife Priority Research Areas under the program Excellence Initiative-Research \mbox{University} at the Jagiellonian University.

\end{multicols*}
\end{document}